\documentclass{article}
\usepackage{frascatiphys}

\begin{document}
\title{ MULTISCALE IMAGE ANALYSIS APPLIED TO $\gamma$/HADRON DISCRIMINATION FOR VERY HIGH ENERGY GAMMA RAY 
        ASTRONOMY WITH ARGO-YBJ }

\author{
I. De Mitri  ,   F. Salamida       \\
{\em Dipartimento di Fisica - Universit\`a di Lecce and INFN, Lecce, Italy} \\
and G.Marsella \\
{\em Dip. di Ing. dell'Innovazione - Universit\`a di Lecce and INFN, Lecce, Italy} \\
\\ On behalf of the ARGO-YBJ Collaboration
}

\maketitle
\baselineskip=11.6pt
\begin{abstract}
Intrinsic differences in the processes involved in the electromagnetic and hadronic shower development 
in the atmosphere have been evidenced by means of a careful analysis of the event image given by the ARGO-YBJ 
detector. The images have been analyzed at different length scales and their multifractal nature has been 
studied. The use of the multiscale approach together with a properly designed and trained Artificial Neural 
Network, allowed us to obtain a good discrimination power. 
If confirmed by further studies on different event cathegories, this result would allow to nearly double the 
detector sensitivity to gamma ray sources.  
\end{abstract}
\baselineskip=14pt

\section{Introduction}
\label{sec:intro}
\begin{figure}
\vspace{6.0 cm}
\includegraphics{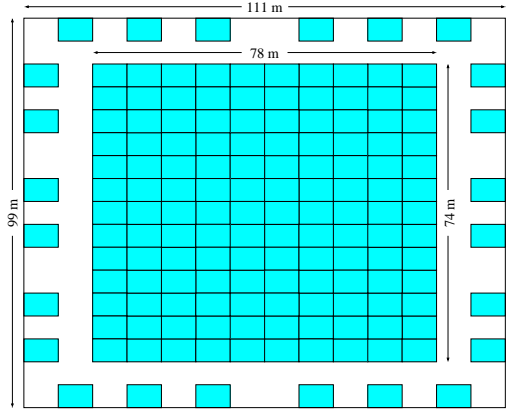}
\vskip -0.5cm
\caption{\it  \label{fig:argo} General layout of the ARGO-YBJ detector. Each indicated box 
represents a  $7.64 \times 5.72 \,$m$^2$ cluster of 12 RPCs (see text). }
\end{figure}

Gamma ray astronomy at energies around 100 $GeV$-10 $TeV$ is the main scientific goal of the ARGO-YBJ 
experiment \cite{argo1,argo2}. 
The detector, which is now being assembled in Tibet (China) at 4300 $m$ a.s.l., is a full coverage Extensive 
Air Shower array consisting of a Resistive Plate Chamber (RPC) carpet of more than 6000 $m^{2}$.
It is logically divided into 154 units called {\it clusters} ($7.64 \times 5.72 \,$ m$^2$), made by 
12 RPCs (see Fig.\ref{fig:argo}).
Each RPC ($1.26 \times 2.85 \,$ m$^2$) is read out by 10 pads ($62 \times 56 \,$ cm$^2$), which are 
further divided into 8 different strips ($62 \times 7 \,$ cm$^2$), which provide the highest available 
space resolution.
The signals coming from all the strips of a given pad are sent to the same channel of a multihit TDC. 
The whole system is designed in order to provide a single hit time resolution at the level of 1 ns, thus 
allowing a complete and detailed three-dimensional reconstruction of the shower front.
The high altitude ( $\sim 606 \,$ g/cm$^2$ ) and the full coverage ensure a very low primary energy 
threshold (E$_{\gamma}\approx$100 GeV), while the detector time resolution 
gives a good pointing accuracy, thus allowing a high sensitivity to $\gamma$-ray sources. 

Gamma/hadron discrimination is a key issue in Very High Energy (VHE) gamma ray astronomy since it allows, 
together with a good angular resolution, the rejection of the huge background due to charged primary hadrons. 
The use of a full coverage detector with a high space granularity - like ARGO-YBJ - can give detailed 
images of the shower front. Intrinsic differences in the processes involved in the electromagnetic and hadronic 
shower development in the atmosphere can then be evidenced by means of a careful analysis of the 
event\cite{miller,busmari}.
Recently the use of multiscale behavior of event images has been showed to give good results in experiments
exploiting the Imaging Atmospheric Cherenkov Techniques (IACT), together with the use of the so-called 
Hillas parameters \cite{tactic,hegra}.
In these experiments the image is integrated over the entire shower development, while in the case of 
ARGO-YBJ a section of the shower is provided at a given (slanted) depth only, thus giving a potential 
lack of information. 
However the typical disuniformities present in hadronic events might be better evidenced in the 
case of ARGO-YBJ , these being partially masked in the case of IACT detectors because of the smearing 
effect due to the integration of the information along the shower development.

In this work event images have been analyzed at different length scales and their multifractal nature 
has been studied. In particular the Discrete Wavelet Transforms have been applied since they allowed a 
differential approach to multifractality, that gave a higher discrimination power.

\section{The multiscale approach}
\label{sec:multiscale}

\subsection{MultiFractal Analysis (MFA)}
\label{sec:MFA}
We have considered the shower image seen by ARGO-YBJ as a function $\phi (\vec{x})$ defined on a 
two-dimensional space and corresponding to the amplitudes given by the measured strip multiplicity. 
As a first step we can calculate the multifractal 
moment $Z_{q}(\ell)$ of order q, defined (see  \cite{dwa}) at the length scale $\ell$ as:
\begin{equation}
\label{eq:zq}
Z_{q}(\ell)=\sum_{\{\vec{x}\}}p(\vec{x},\ell)^{q}
\end{equation}
where
\begin{equation} 
\label{3.mult1}
p(\vec{x},\ell)= \frac{1}{N_{tot}} \int_{B_l(\vec{x})}   \phi (\vec{x}') \, d \vec{x}' 
\end{equation}
gives the probability for a hit to be in the box $B_l(\vec{x})$ of dimension $\ell$ centered at 
$\vec{x}$, being $N_{tot} = \int \phi (\vec{x}') \, d \vec{x}'$ the total content of the image. 
By dividing the image, at different steps, into non overlapping pixels 
of size $\ell$, we can evaluate the box-amplitudes $p(\vec{x},\ell)$. As pointed out in \cite{tactic,hegra,dwa}, 
at each order $q$, the MF moment is expected to have a power law dependence on $\ell$ in the high resolution 
limit, namely: $Z_{q}(\ell) \sim \ell^{\tau(q)}$ for $\ell \rightarrow 1$.

By fitting the behavior of $Z_{q}(\ell)$ on $\ell$ (for each value of $q$) the MF scaling exponent $\tau(q)$ 
can be extracted. The dependence of $\tau(q)$ on 
$q$ gives the main information on the MF properties of the image.

\subsection{The Discrete Wavelet Transform Analysis (DWTA)}
\label{sec:DWTA}
\begin{figure}
\vspace{3.0 cm}
\includegraphics{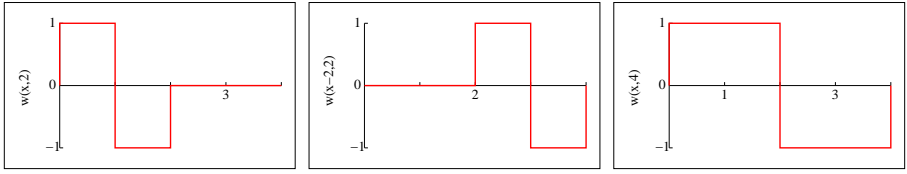}
\vskip -0.7cm
 \caption{\it  Examples of Haar wavelet functions $w(x,2\ell)$ for $\ell$=1,2 \label{fig:haar}}
\end{figure}

As shown in \cite{dwa}, the MF approach might not sufficiently characterize the image.
An approach to multifractality based on the discrete wavelet transformations (DWT) is more appropriate.
DWT can be seen as an expansion of $\phi (\vec{x})$ on a discrete set of basis functions that are 
generated by scaling a so-called mother wavelet. 
For sake of simplicity, let us consider a one-dimensional case. 
If the Haar mother wavelet $w(x,\ell)$ is chosen (see Fig.\ref{fig:haar}), the differences between the box 
amplitudes of two adjacent cells can be given by the proper convolution:
\begin{equation}
\label{eqapp1.15}
p(x,\ell)-p(x+\ell,\ell)= \frac{1}{N_{tot}} \int \phi(x')\;w(x'-x,2\ell) \, dx'
\end{equation}
In this simple case, the mother wavelet can be indicated as (+ $\,$ -).
A differential approach to multifractality is then given by defining the DWT moment as:
\begin{equation}
\label{3.wave}
W_{q}(\ell)=\sum_{\{x\}}|p(x,\ell)-p(x+\ell,\ell)|^{q}
\end{equation}
Also in this case the scaling properties of the image can be evidenced at high resolutions:
$W_{q}(\ell) \sim \ell^{\beta(q)}$ when $\ell \rightarrow 1$.

In the case of a two-dimensional distribution (i.e. an image) we need three base wavelets that can be 
indicated as:
\begin{equation}
\label{eqapp1.26}
\begin{array}{c}
\left( \begin{array}{cc}
+ & + \\
- & -
\end{array} \right)\\
\\(1)
\end{array}
\quad
\begin{array}{c}
\left( \begin{array}{cc}
+ & - \\
+ & -
\end{array} \right)\\
\\(2)
\end{array}
\quad
\begin{array}{c}
\left( \begin{array}{cc}
+ & -\\ 
- & +
\end{array} \right)\\ 
\\(3)
\end{array}
\end{equation}
This led to write three different DWT moments $W^{(i)}_{q}(\ell)$, $i=1,2,3$, defined as:
\begin{eqnarray}
W^{(1)}_{q}(\ell) = \sum_{\{ x,y \}} |p(x,y,\ell)+p(x+\ell,y,\ell) 
                    -p(x,y+\ell,\ell)-p(x+\ell,y+\ell,\ell)|^{q}                \nonumber  \\
\label{eq:wq}
W^{(2)}_{q}(\ell)=\sum_{\{ x ,y \}} |p(x,y,\ell)-p(x+\ell,y,\ell) 
+p(x,y+\ell,\ell)-p(x+\ell,y+\ell,\ell)|^{q}
\\
W^{(3)}_{q}(\ell)=\sum_{\{ x,y \}} |p(x,y,\ell)-p(x+\ell,y,\ell) 
-p(x,y+\ell,\ell)+p(x+\ell,y+\ell,\ell)|^{q} \nonumber
\end{eqnarray} 
These functions are expected to scale like $W^{(i)}_{q}\sim \ell^{\beta^{i}(q)}$, when $\ell \rightarrow 1$,
with different exponents $\beta^{i}(q)$, whose dependence on $q$ gives the maximum information on the 
image properties.

\section{The Monte Carlo Samples}
\label{sec:simulated}
We have generated $\sim$2.8$\cdot$10$^{5}$ gamma-initiated showers and $\sim$2.6$\cdot$10$^{5}$ 
proton-initiated ones making use of the CORSIKA code \cite{corsika}. 
The events have been taken within the energy range 30 GeV$\div$100 TeV with azimuth between 0 and 15 
degrees and core at the detector center. 
The primary energy spectrum has been generated according to the measured power laws with spectral 
index $\gamma$=2.5 for gammas and $\gamma$=2.7 for hadrons.

The detector response has been fully simulated by using ARGOG, a tool 
(based on the GEANT3 package \cite{geant3}) developed within the ARGO-YBJ collaboration.
Since it is the hit multiplicity which is actually measured in EAS experiments, it is correct
and appropriate to classify events following this variable and not in terms of the primary energy.
Therefore we divided the simulated data sample into five multiplicity windows as reported in 
Tab.\ref{tab:ann}, where the average primary energies are also shown. 
The $\gamma$ and proton energies are obviously different since they depend on the processes involved in the 
shower development in the atmosphere. In particular
proton-induced showers results in lower multiplicity events with respect to the case of photon-induced ones 
with the same primary energy.
Moreover in order to avoid distortion effects due to the finite size of the energy window in which the
samples have been simulated, we have considered in the analysis only events which produced more than 50 
and less than 6000 hits. This corresponds to average primary photon energies between 
$E_{\gamma} \sim 500 \,$GeV and $E_{\gamma} \sim 10 \,$TeV (see Tab.\ref{tab:ann}).

\section{Event analysis}
\label{sec:event}
The multiresolution quantities defined in Sec.\ref{sec:multiscale} have been used to analyze each event image.
As reported in Sec.\ref{sec:intro} the ARGO-YBJ detector is made by a central carpet and a guard ring
(see Fig.\ref{fig:argo}). In order to preserve the same symmetry at different lenght scales $\ell$, 
we decided to neglect, in this first analysis, the information coming from the external ring. 
We also decided to {\it mask} the central carpet with a square grid. 
In particular, since it is made of (120$\times$130)pads, the first and the last row of pads were not 
considered, while four empty columns of pads were added (two on the left and two on the right), 
thus obtaining a (128$\times$128) pad mask.
In order to limit statistical fluctuations of the hit multiplicity in the smallest pixels,
the minimum pixel size considered in the analysis was set at (2$\times$2)pads - about $1.4 \,$m$^2$ - which 
will then corresponds to the maximum resolution, i.e. $\ell$=1.  
The analysis of an event goes then throught different steps, 
each correponding to different lenght scales. At the $n$-th step (with $n=1,2,...,6$), the image is divided 
into 2$^{n} \times$2$^{n}$ square pixels of size $ \ell_{n} = \frac{64}{2^{n}} $, containing each 
$4 \ell_{n}^2$ pads, and the total strip multiplicity is computed in each considered pixel.

\begin{figure}
\begin{tabular}{cc}
\includegraphics{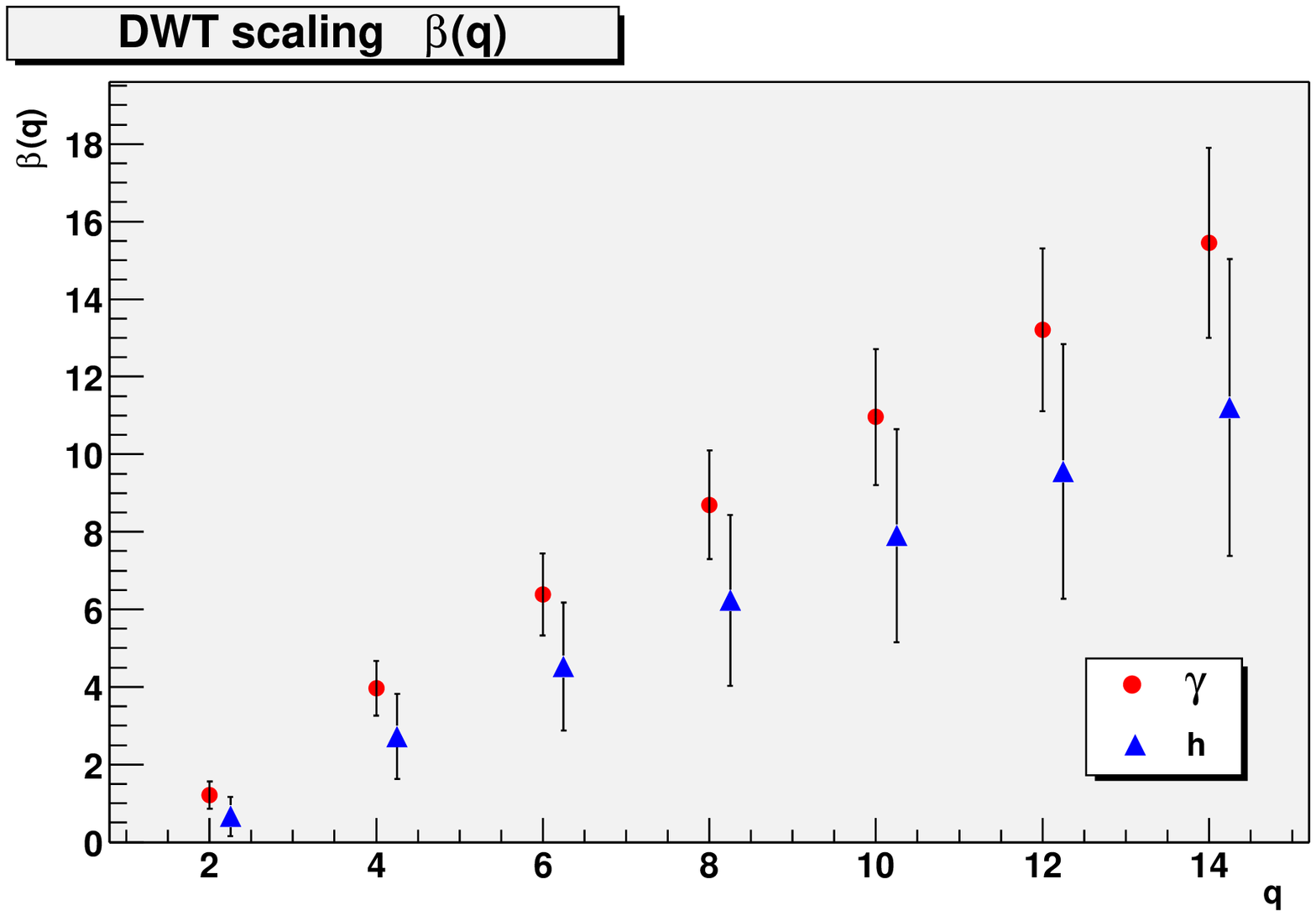}
&
\includegraphics{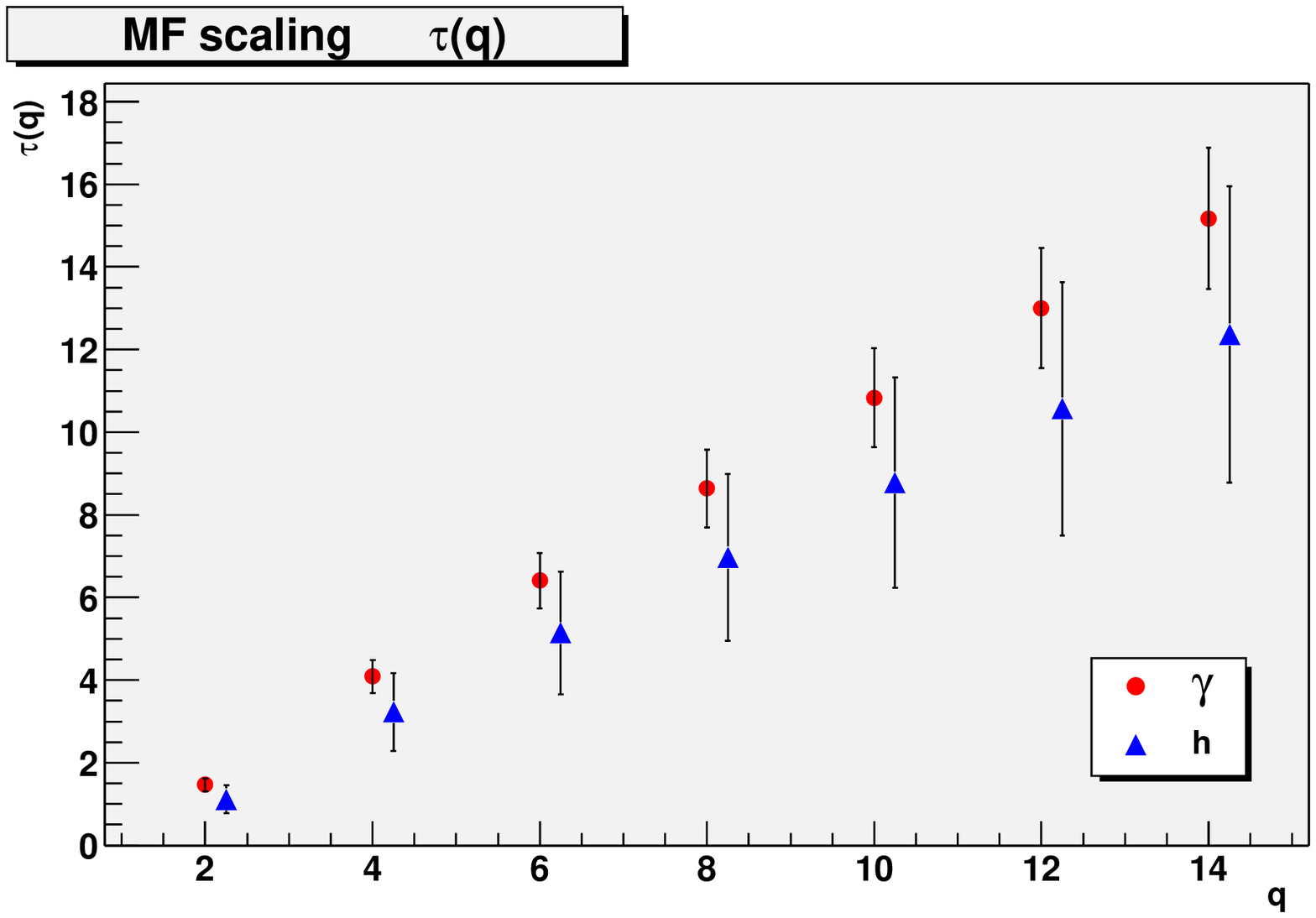}
\end{tabular}
\vskip 4.cm
\caption{\it Dependences of the scaling exponents on the moment $q$ for a sample of gamma and proton initiated-showers 
with energy from 3 $TeV$ to 10 $TeV$ . The error bars refer to the r.m.s values of the variuos distributions.
\label{fig:scalingvsq} }
\end{figure}

The values of Log(Z$_q$($\ell$)) and Log(W$_q$($\ell$)) have been calculated following 
Eq.\ref{eq:zq} and Eq.\ref{eq:wq}, and their dependences on Log($\ell$) have been fitted
with a first order polynomial in the region where the scaling is expected (i.e. $\ell \rightarrow 1$ ), 
for different values of the moment $q$. 
The scaling exponents $\tau$(q) and $\beta$(q), defined in Sec.\ref{sec:multiscale}, have then been
obtained for each event.
Their dependences on the moment $q$ are shown in Fig.\ref{fig:scalingvsq} for a sample of gamma 
and proton initiated events. As can be seen there is a separation of the 
average values of the scaling exponents between electromagnetic and hadronic showers, which is however 
partly masked by the large fluctuations. Therefore the separations of the MF and DWT parameters 
are not sufficient in order to give a good discrimination between e.m. and hadronic showers, unless an 
Artificial Neural Network (ANN) is used as in ref.\cite{tactic,hegra}. 
We then decided to use multifractal parameters as inputs to a properly
designed and trained ANN (see Sec.\ref{sec:ann}).

\begin{figure}[h!]
\begin{tabular}{cc}
\includegraphics{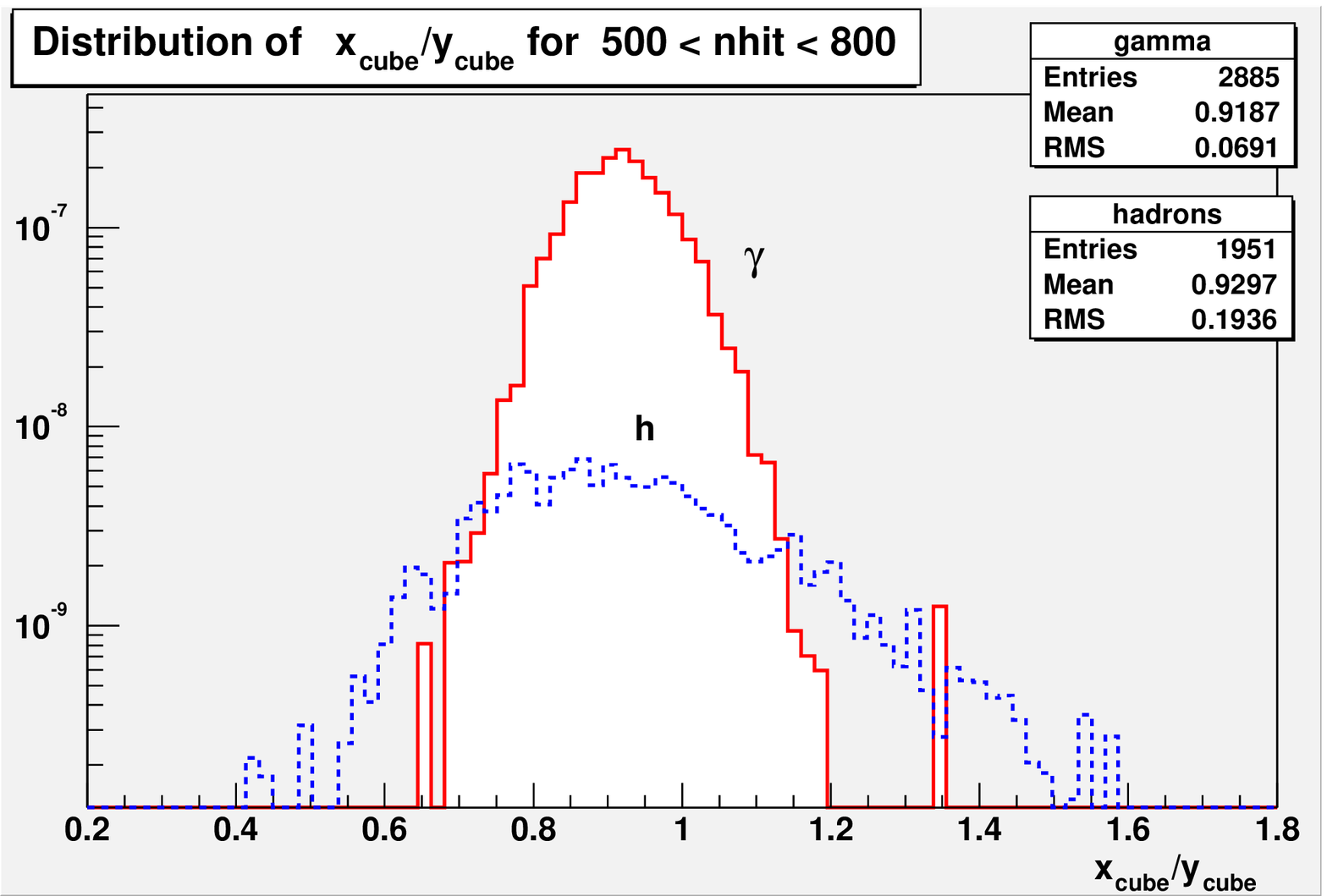}
&
\includegraphics{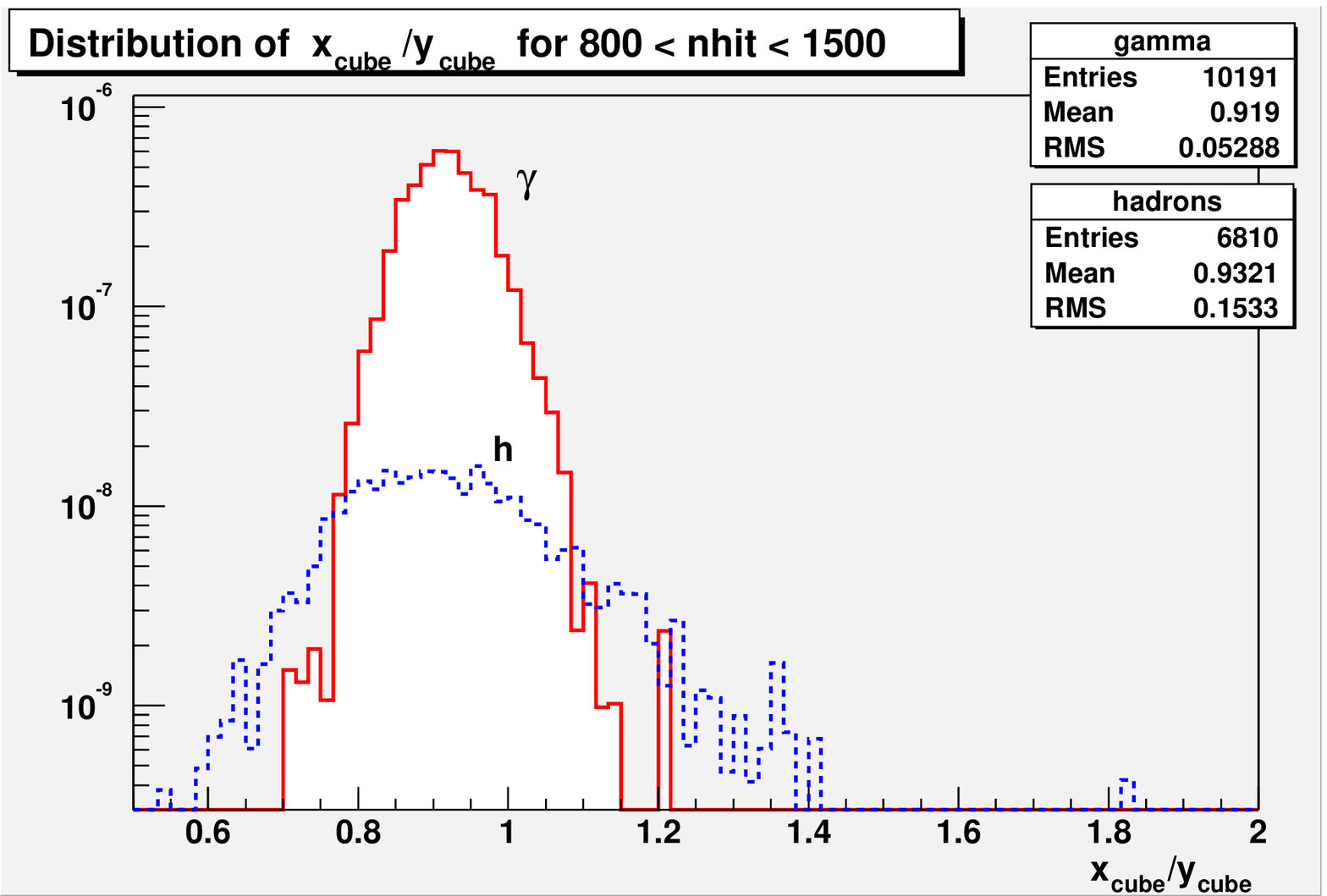}
\end{tabular}
\vskip 4.cm
\caption{\it Distributions of $\xi$ for simulated events in a couple of multiplicity regions (see text).\label{fig:xi}}
\end{figure}

In order to increase the $\gamma$/h separation, a study on the shape and the simmetry of the event image has also
been made. In particular we studied the skewness of each event by means of the third moment of the 
distributions of the hit coordinates on the detector plane, namely $x$ and $y$. In general, the skewness of
a sequence of values ${\cal S} \equiv {x_1, x_2, ..., x_1, ..., x_N}$ is defined as:
\begin{equation}
\label{skw}
\eta = \frac{ \sum_{i=1}^{N}\left( x_{i} - \overline{x} \right)^3 }{ \left( N-1 \right) s ^{3}}
\end{equation}
where $\overline{x}$ and $s$ are the average and r.m.s. over ${\cal S}$, respectively. 
In our case we studied the behaviour of the quantity $\xi = x_{cube}/y_{cube}$, with:
\begin{equation}
\label{m3}
x_{cube}=\frac{\sum_{i}n_{i}x_{i}^{3}}{\sum_{i}n_{i}}
~~~~~~~~~~,~~~~~~~~~~
y_{cube}=\frac{\sum_{i}n_{i}y_{i}^{3}}{\sum_{i}n_{i}}
\end{equation}
where $n_i$ is the strip multiplicity of the pad at the position $(x_i,y_i)$. 
The value of $\xi$ is reported in Fig.\ref{fig:xi} for two samples of e.m. and hadronic showers in two different
multiplicity windows. As can be seen, its average is almost one as expected from both its definition and 
the detector geometry, while the r.m.s is always larger for proton-initiated showers, due to the large
fluctuations present in hadronic events. This behaviour suggested the use of $\xi$ as one of the inputs to 
the ANN.

\section{The Artificial Neural Network}
\label{sec:ann}
In order to perform the $\gamma$/hadron discrimination using the parameters we have introduced in the 
previous sections, we decided to make use of an Artificial Neural Network. The Neural Network we have 
chosen is of the {\it feed forward} type and it is made of 3 perceptrons layers. The ANN input is an 
eight-dimensional vector whose elements are:
the event total hit multiplicity $N_{hit}$, 
the value of $\xi$, 
the multifractal exponents $\tau$ and $\beta$ for $q=4,6,8$.
The output vector is defined in a one dimensional space: it is trained to be 1 for gamma-initiated events
and 0 for hadronic showers.

\begin{figure}[h]
\begin{tabular}{cc}
\includegraphics{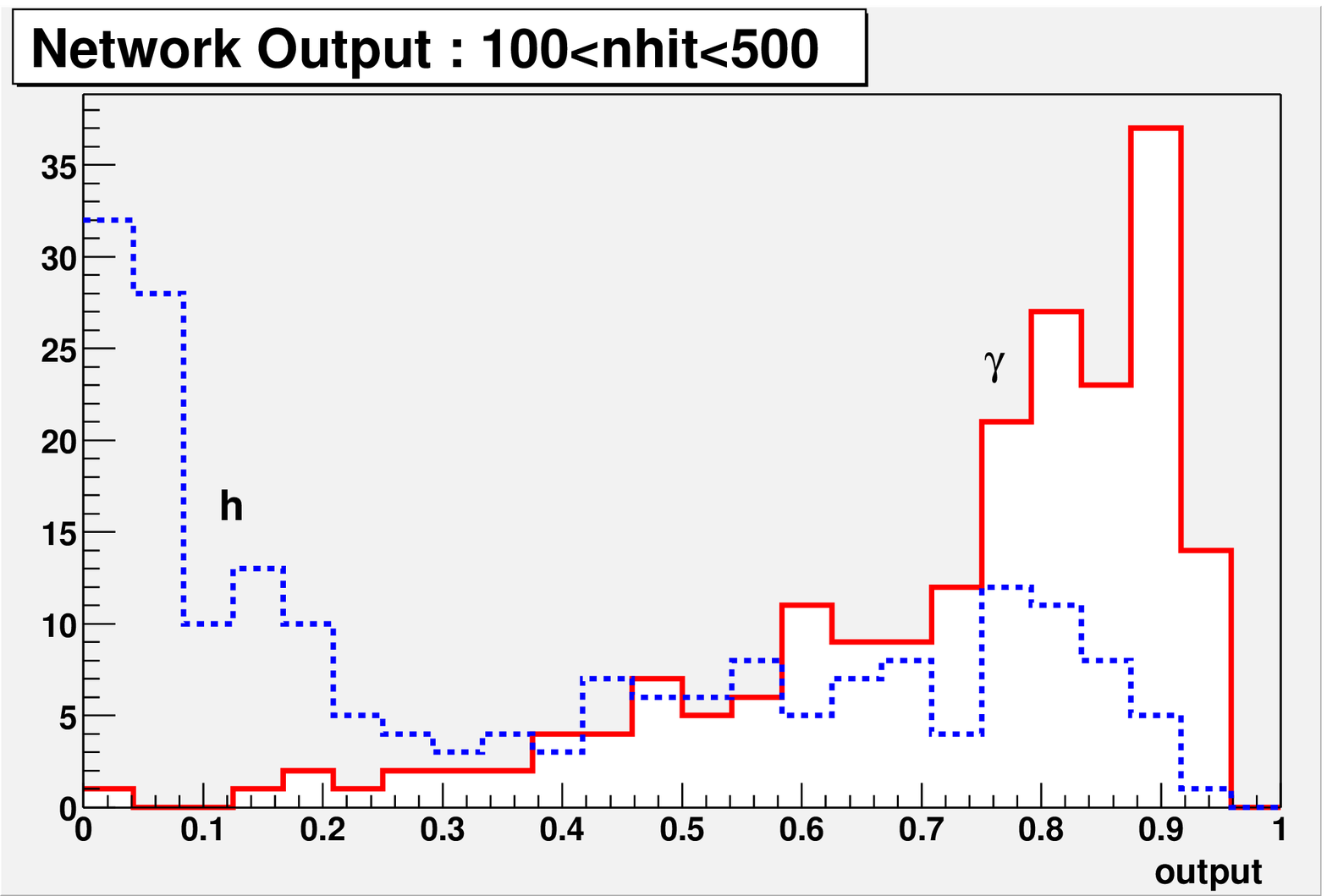}
&
\includegraphics{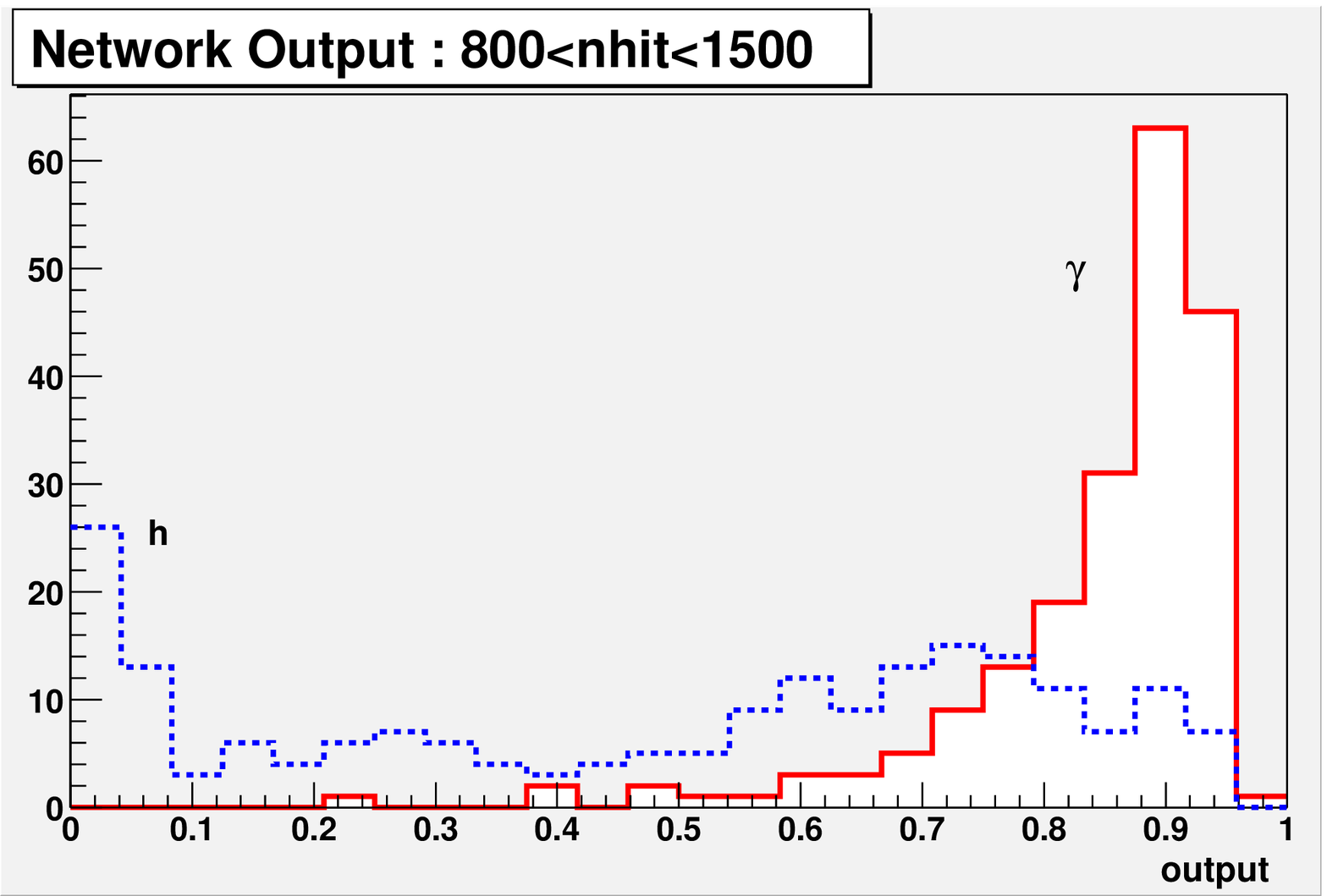} 
\end{tabular}
\vskip 5.cm
\caption{\it Outputs of the neural network in two of the five considered multiplicity regions. \label{fig:ann}}
\end{figure}

Networks were implemented and optimized by using the {\bf S}tuttgart {\bf N}eural {\bf N}etwork {\bf S}imulator
(SNNS) tool \cite{snns}. 
In designing the ANN, its characteristics were been deeply studied in order to reach a good 
compromise between the increase of recognition capability and the processing time. 
The network training was separately performed in 5 multiplicity windows by using several thousands events 
(see Tab.\ref{tab:ann}).
The ANN were then tested by using an independent reduced sample of events and the $\gamma$ recognition 
efficiency $\varepsilon_\gamma$, together with the proton contamination $(1-\varepsilon_{p})$,
were measured. The quantities $\varepsilon_\gamma$ and $\varepsilon_p$ reached their plateau value after 
few thousands of ANN training epochs.
An example of ANN output for a couple of multiplicity windows is shown in Fig.\ref{fig:ann}.

The detector sensitivity to $\gamma$-ray sources is defined as $S = N_{\gamma}$/$\sqrt{N_h}$, where 
$N_{\gamma}$ is the number of gamma-initiated events, while $N_h$ is the hadron contamination of the
considered sample. The use of a $\gamma$/h discrimination tool like the one we are considering in this work
makes the sensitivity $S$ to be multiplied by the factor $Q=\varepsilon_{\gamma}$/$\sqrt{1-\varepsilon_{h}}$.

In this work values of $Q\approx$2 have been reached (see Tab.\ref{tab:ann}), which are among the largest 
obtained with various techniques in the experiments working in the field \cite{hegra}. 
This result would allow to nearly double the ARGO-YBJ sensitivity to a given source obtained from the
pointing accuracy alone\cite{vernettoicrc} or, equivalently, to reduce by a 
factor four the time needed to observe it above the hadron background, with a given statistical 
significance.

\begin{table}[t]
\centering
\begin{tabular}{|c|c|c|c|c|c|}\hline
nhits & events ($\gamma$) & events (p) & $<$E$_p>$  & $<$E$_{\gamma}>$ & Q  \\ \hline \hline
50$\div$100    & 6657  & 3862 & 0.8 TeV  & 0.5 TeV    & 1.28$\pm$0.01 \\  \hline
100$\div$500   & 11556 & 6862 & 1.8 TeV  & 1.1 TeV    & 1.42$\pm$0.02 \\  \hline
500$\div$800   & 2571  & 1644 & 4.9 TeV  & 2.9 TeV    & 2.01$\pm$0.10 \\  \hline
800$\div$1500  & 3087  & 1963 & 7.6 TeV  & 4.6 TeV    & 1.78$\pm$0.07 \\  \hline
1500$\div$6000 & 4329  & 3053 & 18.4 TeV & 11.3 TeV   & 1.78$\pm$0.06 \\  \hline
\end{tabular} 
\linespread{1.0}
\caption{\label{tab:ann}\it Main characteristics of the simulated data sample 
(no. of ANN training events, average primary energy, ....) together with the values of 
$Q$ for $\gamma$/h discrimination that resulted from this work.}
\end{table}

\section{Concluding remarks}

We presented the first results of a multiscale image analysis performed on Monte Carlo events by taking into 
account all the processes of shower development in the atmosphere and a full simulation of the ARGO-YBJ 
detector response. The images have been analyzed at different length scales and their multifractal 
nature has been studied. 
A set of eight image parameters has been identified and used as the input for an Artificial Neural Network, 
which was then trained in order to discriminate gamma initiated from proton-initiated showers.

Since this is the first attempt of this kind of analysis in a EAS detector like ARGO-YBJ , we decided to 
restrict this study to events with the core at the detector center and azimuth angles not larger than 15 
degrees, while all the energies with the correct spectral dependencies have been simulated in a wide 
range. At this level we also neglected the contribution given by primary nuclei heavier than protons.
This is a good first order approximation because of the proton-dominated cosmic ray composition in the
considered energy region. Furthermore heavier-nuclei-induced showers would produce event patterns with 
characteristics even more different from gamma-initiated ones. 

If the results obtained in this first study will be confirmed by a further analysis on the whole event 
categories (now in progress), the detector sensitivity to a given source would 
nearly double or, equivalently, the time needed to observe it above the hadron background, with a given 
statistical significance, would be reduced by a factor four.
 
The best performances in $\gamma$/h discrimination have been obtained for photon primary energies 
in the few TeV range, while at higher energy this analysis might be well complemented by measuring the
muon content of the shower \cite{fratini}.


%

\begin{thebibliography}{99}
\bibitem{argo1} C.Bacci et al., Nucl. Instr. \& Meth. in Phys. Res. {\bf A443}, 342 (2000)
\bibitem{argo2} C.Bacci et al., Astroparticle Phys. {\bf 17}, 151 (2002) 
\bibitem{miller} R.S.Miller and S.Westerhoff, Astroparticle Phys. {\bf 11}, 379 (1999)
\bibitem{busmari} S.Bussino and S.M.Mari, Astroparticle Phys. {\bf 15}, 65 (2001)
\bibitem{tactic} A. Haungs et al., Astroparticle Phys. {\bf 12} 145 (1999)
\bibitem{hegra}B. M. Sch$\ddot{a}$fer et al., Nucl. Instr. $\&$ Meth. in Phys. Res., {\bf A465} 342 (2001)
\bibitem{dwa}Jan W. Kantelhardth et al., Physica {\bf A220} 219 (1995)
\bibitem{corsika} D.Heck et al., Report FZKA 6019, Forschungszentrum, Karlsruhe (1998)
\bibitem{geant3} R.Brun et al., CERN Publication DD/EE/84/1 (1992)
\bibitem{snns}{\it http://www-ra.informatik.uni-tubingen.de/SNNS/}
\bibitem{vernettoicrc} S.Vernetto et al., Proceedings of the 28$^{th}$ ICRC Conference, Tsukuba, 2003.
\bibitem{fratini} See K.Fratini et al., these proceedings.
\end{thebibliography}
\end{document}